\documentclass[12pt]{article}
\usepackage[utf8]{inputenc}
\usepackage{graphicx}
\usepackage{float}
\usepackage{indentfirst}
\usepackage{cite}
\usepackage{gensymb}
\usepackage[margin=1.0in]{geometry}
\usepackage{setspace} 
\usepackage{tabularx}
\usepackage{wrapfig}
\usepackage[labelfont=bf]{caption}
\usepackage[font={footnotesize}]{caption}
\usepackage[nottoc]{tocbibind}
\usepackage{hyperref}
\usepackage{authblk}
\usepackage{amsmath}
\usepackage[table, dvipsnames]{xcolor}
\usepackage{bmpsize}

\hypersetup{
    colorlinks,
    citecolor=black,
    filecolor=black,
    linkcolor=black,
    urlcolor=black
}
\usepackage{subfig}
\begin{document}

\title{Unconstrained dynamic gel swelling generates transient surface deformations}

\author[1]{Alyssa VanZanten}
\author[2]{Shih-Yuan Chen}
\author[2]{Michelle M.\ Driscoll}
\author[1]{Caroline R.\ Szczepanski}
\affil[1]{Department of Chemical Engineering and Materials Science, Michigan State University}
\affil[2]{Department of Physics and Astronomy, Northwestern University}

\maketitle

\section*{Abstract}

Polymer gels are comprised of a three-dimensional, cross-linked network that can typically withstand the mechanical deformation associated with both swelling and de-swelling. Thus, gels can be designed with smart behaviors that require both stress generation and dissipation, making them well-suited to many applications including membrane technology, water capture devices, and drug delivery systems. In contrast to the fully swelled equilibrium state, limited research characterizes the unsteady-state swelling regime prior to equilibrium. It is in this regime where unique surface deformations can occur. Here we show how internal network constraints and external diffusive pressure can be leveraged to manipulate swelling kinetics and surface deformations in poly(ethylene glycol) gels during unconstrained, three-dimensional swelling. We find that increasing cross-linker molecular weight and swelling in ethanol, as opposed to water, are both effective routes to increase the time it takes to reach equilibrium but do so through different mechanisms. Networks with fewer internal constraints, manipulated via cross-linker chain-length, imbibe more solvent over a longer time. In contrast, swelling in ethanol reduces the amount of solvent imbibed by the network while increasing the time to reach equilibrium. Measurements of surface patterns during swelling establishes that an immediate, fast relaxation at the surface occurs during the first five minutes of swelling. However, the density and persistence of these features varies with solvent quality. These results serve establish a framework for how soft materials undergo dynamic deformation. Engineering transient surface properties while mitigating unwanted instabilities opens the door for emerging technologies such as smart anti-fouling and sensors.

\newpage

\doublespacing
\section{\textbf{Introduction
}}
The development of smart soft materials in recent years has largely focused on utilizing stimuli-responsive polymers, because of the numerous synthetic pathways available to tailor their responsive behavior \cite{Xin2010, Zeng2017, Huang2017, Manouras2017}.
For example, hydrogels, which are cross-linked, elastic polymer networks capable of absorbing large amounts of liquid, can be designed to swell and de-swell in response to external stimuli, like heat, solvent quality, and light \cite{Jiang2020 ,Wang2016}. These qualities make hydrogels attractive for targeted drug delivery systems \cite{Brazel, Colombo1994} and tissue engineering \cite{Devolder2012, Gargava2016}.
For instance, controlled release in delivery applications requires precise tailoring of transport kinetics \cite{Saidi2020} and internal forces \cite{Tanaka1992}, such as the generation and dissipation of stresses during gel swelling.
Interestingly, the significant internal stresses that arise during swelling are known to induce instabilities, such as bulk buckling and surface deformation \cite{Takahashi2016, Weiss2013, Kim2014}.
These instabilities are relevant and must be understood at a fundamental level, since changes at the surface of a material can dramatically impact its interactions with surrounding fluids.
Thus, establishing how synthetic handles, such as cross-link density, impact surface deformation is critical when tailoring gels for specific application fields.
This work utilizes hydrogel swelling as a platform to establish the fundamental stress generation and dissipation mechanics that lead to reversible surface deformation in soft materials.

To understand swelling of a polymer network at a fundamental level, researchers often turn to Flory-Rehner polymer solution theory, which posits that gel swelling is driven by two opposing forces: (1) the force of mixing that results from the attraction between the polymer network and the solvent, and (2) the opposing elastic force that is due to the resistance of the elastic network to stretching during swelling \cite{Flory1943, Flory1943swelling, Louf2021}. This theory successfully summarizes the complex interplay of forces that govern swelling, and therefore captures a broad range of gel swelling phenomena, such as the existence of continuous and discontinuous volume phase changes, and the effects of temperature, pH, and salt concentration on swelling \cite{Quesada-Perez2011}.
However, while Flory-Rehner describes the fully swelled, equilibrium state, there are systems to which this theory cannot be applied. For example, the use of macromeric (e.g., longer) cross-linkers, which are common in hydrogel development, deviates from the small-chain cross-links proposed by Flory-Rehner\cite{Borges2020, Borges2023}. Furthermore, Flory-Rehner assumes that a polymer network is in an equilibrium state (i.e., that the chemical potential ($\mu$) is equivalent inside and outside of the gel); When considering the unsteady-state swelling regime prior to equilibrium, this assumption no longer holds true. Thus, understanding transient swelling behavior prior to an equilibrium state remains a challenge. 
Bulk instabilities such as spontaneous fracture and dynamic surface deformation arise during non-equilibrium swelling, which highlights the need to understand this transient behavior \cite{Leslie2021, Takahashi2016}. Understanding the origin of these instabilities will provide information on how localized stress gradients arise during swelling, as well as the mechanisms by which mismatched internal stresses are dissipated during transient swelling. 

Despite the limited characterization of transient swelling, prior studies on equilibrium systems provide guidance as to what parameters can significantly impact gel swelling. 
For instance, we know that equilibrium swelling depends on cross-linking \cite{Peppas1977}, polymer-solvent interaction \cite{Punter2020}, and environmental conditions \cite{Louf2021, Molina2007}. Additionally, studies that analyze the kinetics of swelling show that geometry \cite{Li1990}, polymer-solvent interaction \cite{Drozdov2016, Rehner1975}, and diffusion rate \cite{Tanaka1979} strongly impact swelling, but more work is needed to understand how these factors influence the dynamics of internal stress during the entire swelling process.
Building upon this foundation, here we examine the changes in transient swelling and instability behaviors (e.g. surface deformations such as wrinkling, creasing, and folding) due to variations in both network constraints and solvent quality.

Many models of hydrogel swelling have been developed to capture these instability behaviors, which are driven both by solvent diffusion and non-linear, viscoelastic deformation of the polymer network \cite{Chester2012, Drozdov2016, Plummer2023, Curatolo2017, Guo2020, Islam2021, Yang2021}. 
However, these models typically assume
a two-dimensional material geometry consisting of a thin `skin' layer adhered to a softer foundation. 
With this geometry, 
the wavelength and amplitude of wrinkles on the surface are related to the compressive force in the skin layer \cite{Genzer2006}. 
Furthermore, in many two-dimensional systems, surface deformations are engineered by imposing bulk strain, either manually through stretching \cite{Yang2010, Cerda2003, Yang2010} or by constraining a gel to a substrate during swelling \cite{Tanaka1992, Trujillo2008, Kim2014, Yang2010, Ju2022, Xu2013, Dupont2010,Toh2015,Ju2022}, as well as by creating a depth-wise cross-linking gradient \cite{Guvendiren2009, Kim2014, Gu2017, Guvendiren2010}.
While these models and prior works characterize and describe behaviors such as wrinkling in coatings and thin films, 
assumptions that the elastic modulus and skin layer thickness are at steady-state do not apply to the
system presented herein, which features
free, unconstrained gel swelling, where material parameters are transient.
While characterizing surface deformations in a transient, unconstrained system is challenging, it is crucial
for engineering applications as these instabilities can be leveraged to impart anti-fouling or adhesive properties \cite{Genzer2006, Rodriguez-Hernandez2015}.

To describe the network response during swelling, we can consider three aspects that evolve dynamically and are inter-related:
(1) swelling kinetics, (2) stress dynamics, and (3) surface deformation mechanics. \textit{Swelling kinetics} defines the rate at which solvent is transported into and through a gel, and is dependent on many factors, such as cross-link density and polymer-solvent attraction \cite{Li1990, Tanaka1979, Scholt1992, Guvendiren2010, Saidi2020}. Swelling kinetics depend on how effectively polymer chains can move and stretch in response to swelling, as well as diffusion of solvent molecules. When swelling begins, the surface of a gel becomes saturated and stretches to accommodate diffusing solvent while being constrained by the inner portion of the material that is not yet expanding or responding to swelling, since the solvent molecules have not yet diffused to the interior. These kinetics broadly influence \textit{stress dynamics}, such as the stress that builds up due to mismatched swelling in the outer versus inner regions and the compressive stress experienced at the surface of the gel \cite{Brazel}. The magnitude of stress experienced locally changes based on the speed at which the saturated outer region diffuses inwards toward the unsaturated center.
As a result of internal stress changing and dissipating, \textit{surface deformation} is observed to be transient as swelling progresses.
The solvent-saturated outer layer becomes softer than the bulk \cite{Flory1943swelling, Toh2015}, and this mismatch between layers results in a buildup of compressive stress at the surface, thus leading to deformation (e.g., creasing) \cite{Ilseng2019, Trujillo2008, Weiss2013, Ju2022}. 
The evolution of surface instabilities is directly linked to swelling kinetics, stress dynamics, and therefore network relaxation. Measurements of how creasing patterns evolve during swelling can give us a window into the mechanics of dynamic deformation.

The work presented here utilizes a three-dimensional bulk swelling setup, where the samples are not constrained during swelling and therefore imbibe water on all faces. 
We present unique, transient creasing of poly(ethylene glycol)-based gels and characterize both the swelling kinetics and surface deformation evolution.
In Section 3.1 we characterize the instability behaviors that occur during transient swelling, focusing on the three-stages of surface deformation. We highlight a grid-like surface pattern that emerges after 20 minutes of swelling, which has only been described in a limited context prior to this work \cite{Takahashi2016}.
The impact of network constraints on swelling kinetics and capacity is investigated in Section 3.2 by manipulating cross-link fraction and molecular weight of the cross-linker employed.
We also characterize how external diffusive pressure can be modulated via (co)solvent quality in Section 3.3.
Specifically, swelling in ethanol as opposed to water is shown to decrease the rate of solvent imbibement, while the rate of swelling in water-ethanol (co)solvents changes non-monotonically with respect to ethanol content.
Finally, crease evolution and kinetics are quantified through imaging measurements to infer stress dynamics in Section 3.4.
This study will be used to understand the fundamental factors, both internal and external, that drive network relaxation during swelling.

\newpage
\newpage
\section{\textbf{Experimental Section
}}
\subsection{Materials}
    The base monomer employed in this study is poly(ethylene glycol) methyl ether acrylate (PEGMA) with a molecular weight of 400 g/mol. To form a network structure (gel), a  cross-linker was incorporated into the polymerization. Two different poly(ethylene glycol) diacrylate (PEGDA) cross-linkers were employed, one with an average molecular weight of 700, and the other 10,000 g/mol.  In all formulations, the photoinitiater used was 2,2-dimethoxy-2-phenylacetophenone (DMPA). All materials were obtained from Sigma-Aldrich and were used as received. For swelling experiments, deionized (DI) water was obtained via an in-house DI water source and 200-proof ethanol was obtained from Koptec. (Co)solvent solutions were prepared by mixing DI water and 200-proof ethanol at the following volume ratios: 90:10, 75:25, 50:50, 25:75, and 10:90 (water:ethanol).
\subsection{Methods}
    \subsubsection*{Formulation Preparation and UV Curing}
        To manipulate network structure, two different control parameters were employed when formulating resins: (1) the relative fraction of cross-linker (PEGDA) and (2) the molecular weight of cross-linker employed (PEGDA-700 or PEGDA-10,000), as shown in \textbf{Table \ref{table: formulations}}. Throughout the remainder of the manuscript, formulations will be referenced by their name, which follows the pattern PEGDA(mol\%):PEGMA(mol\%). If the formulation was made with PEGDA-10,000, it is noted by the use of ``10k" at the beginning of the name. If this notation is absent, PEGDA-700 was used in the formulation.
        \begin{table}[H]
        \centering
        \caption{\textbf{Hydrogel formulations and naming scheme}. The hydrogel formulations explored in this study were prepared with different loadings and molecular weights of cross-linker. Rows highlighted pink indicate formulations that typically exhibit surface deformation during transient swelling, and rows highlighted light gray do not exhibit surface deformation but are likely to spontaneously fracture during swelling.}
        \begin{tabular}{|c|c|c|}
            \hline
             & 700 Mn &  \\
            Name & PEGDA (mol\%) & PEGMA (mol\%) \\
            \hline \hline
            \rowcolor{pink} 1:99 & 1 & 99 \\
            \hline
            \rowcolor{pink} 5:95 & 5 & 95 \\
            \hline
            \rowcolor{lightgray} 10:90 & 10 & 90 \\
            \hline
            \rowcolor{lightgray} 20:80 & 20 & 80 \\
            \hline
            \rowcolor{lightgray} 40:60 & 40 & 60 \\
            \hline \hline
             & 10,000 Mn & \\
            Name &  PEGDA (mol\%) & PEGMA (mol\%) \\
            \hline \hline
             \rowcolor{pink} 10k 1:99 & 1 & 99 \\
             \hline
            \rowcolor{pink} 10k 10:90 & 10 & 90 \\
             \hline
            \rowcolor{pink} 10k 20:80 & 20 & 80 \\
            \hline
             \rowcolor{pink} 10k 40:60 & 40 & 60 \\
            \hline
        \end{tabular}
        \label{table: formulations}
        \end{table}
The photoinitiator loading in all formulations was 0.5 wt\%, meaning that the loadings listed for PEGMA and PEGDA in \textbf{Table \ref{table: formulations}} constitute 99.5 wt\% of the total mass.
Resins made with the 700 g/mol PEGDA were prepared neat (i.e., without solvent). In brief, the appropriate mass of DMPA, PEGMA, and PEGDA-700 were incorporated into a glass vial and magnetically stirred until homogeneous ($\sim$ 15 mins). To achieve a homogeneous resin formulation when employing the high molecular weight cross-linker (PEGDA-10K), water was included in the resin. This was incorporated  to homogenize all resin constituents during mixing.
After achieving a homogeneous mixture, the resin was dried at 65\degree C overnight, resulting in at least 95\% of the water being removed as evidenced by the mass change after drying.
    
All polymerized samples were prepared in glass molds with approximate dimensions of 25 x 10 x 2 mm (\textit{l} $x$ \textit{w} $x$ \textit{t}). 
Samples were photocured at an intensity
of 0.1 \(\frac{W}{cm^2}\) in the UV-A region ($\lambda$ $\sim$ 315 to 400 nm). UV irraditation intensity was verified with a UV radiometer (Power Puck II, Electronic Instrumentation and Technology, Inc.).

\subsubsection*{Swelling Investigations}
To measure swelling ratio ($Q$, \textbf{Eq.\ref{eq: Q}}), the mass of the as-prepared, dry, unswollen gel ($m_0$) was recorded prior to the gel being submerged in a large excess (approximately 50 mL) of solvent (deionized water, 200-proof ethanol, or a mixture of both).
The gel was then removed from the solvent at predetermined swelling times. Upon removal, residual solvent was gently dabbed off the sample with a kimwipe, and the mass of the gel ($m(t)$) was recorded. The gel was then placed back into the solvent to continue swelling. This process was repeated over 24 hours of swelling to capture the entire range of behavior, including the equilibrium mass of the gel ($m_{eq}$). For samples that catastrophically ruptured during swelling, all pieces were removed from the solvent, dried, and weighed at each subsequent predetermined swelling time. 
        \begin{equation} \label{eq: Q}
            Q(t)=\frac{m(t)}{m_0},\quad Q_{eq}=\frac{m_{eq}}{m_0}, \quad
            Q_0=\frac{m_0}{m_0}=1
        \end{equation}

Swelling data was collected in triplicate for every formulation (\textbf{Table \ref{table: formulations}}) and solvent condition.
The average and standard deviation of the swelling ratio at a each time point ($t$) were used to calculate $\overline{Q(t)}$, the normalized swelling ratio (\textbf{Eq. \ref{eq: normQ}}) and its associated error.

        \begin{equation} \label{eq: normQ}
            \overline{Q(t)}=\frac{m(t) - m_0}{m_{eq} - m_0}=\frac{Q(t) - Q_0}{Q_{eq} - Q_0} =\frac{Q(t) - 1}{Q_{eq} - 1}
        \end{equation}
In this equation, the relative swelling ratio increase at time ($t$) is divided by the total relative swelling ratio increase at equilibrium. Therefore, the normalized swelling ratio ($\overline{Q(t)}$) represents the fraction of the total swelling capacity that is reached at each time point.

Lastly, initial swelling kinetics were characterized via the swelling rate (e.g., $Q$ as a function of time) during the first 30 minutes of solvent exposure in order to describe the entire range of swelling during which surface deformation (i.e., creasing) is observed.
To obtain this rate, a linear equation was fit to the $Q$ data during the first 30 minutes of swelling. The y-intercept of the linear fit was fixed at 1 as $Q_{0}$ =1 (see, \textbf{Eqn. \ref{eq: Q}}). These initial swelling rate calculations are shown in \textbf{Fig. S\ref{fig: rate in water and ethanol}}.

 
    
\subsubsection*{Hydrogel Morphology Observations and Measurements}
To visualize macroscopic surface deformation during swelling experiments, gel samples were removed, blotted dry, and photographed via iPhone XR. For smaller scale, frequent imaging, \textit{in situ}, optical microscopy (Olympus Microscope ix83) was employed. 
1:99, PEGDA-700 samples were swelled while microscope images (with a field of view of 6656 $\mu$m $\times$ 6656 $\mu$m)
were captured every 20 seconds during the first 31.6 minutes of swelling. To ensure that there was no contact with the surrounding vessel inhibiting diffusion, each hydrogel sample was suspended off the bottom of the dish during swelling. The plane of focus of the microscope was aligned with the bottom of the sample, meaning that creases on the bottom appeared as thin, dark lines, whereas the out-of-focus creases on the top surface show up as larger, dark splotches. See \textbf{Figs. S\ref{fig:crease highlights water}} - \textbf{ S\ref{fig:crease highlights ethanol}} for a clear depiction of the surface features we measured, in which the crease tracings in orange are overlaid on the microscope images.

To extract quantitative data from these microscope images, the crease lines were manually traced using MS Paint on a Microsoft Surface tablet. This process binarizes the images and provides accurate locations for the crease lines. Using a custom Python code, the geometric mean point of each line was identified from the images. Then, the closest neighboring point was identified, any duplicate reported distances were eliminated, and the average distance between neighboring creases was calculated. Once an ensemble of the distance between each crease line was established, the ensemble was averaged to obtain the characteristic ``wavelength" or distance between observed crease patterns.
Three samples were imaged for each solvent mixture. Given the complexity associated with the tracing process, seven equally spaced time points (20, 320, 640, 940, 1300, 1600, and 1900 seconds) were chosen for analysis to capture the range of instability behaviors. 

\newpage
\section{\textbf{Results and Discussion
}}
\subsection{\textbf{Instability Behaviors During Transient Swelling}}
\begin{figure}[ht]
    \centering
    \includegraphics[scale=0.65]{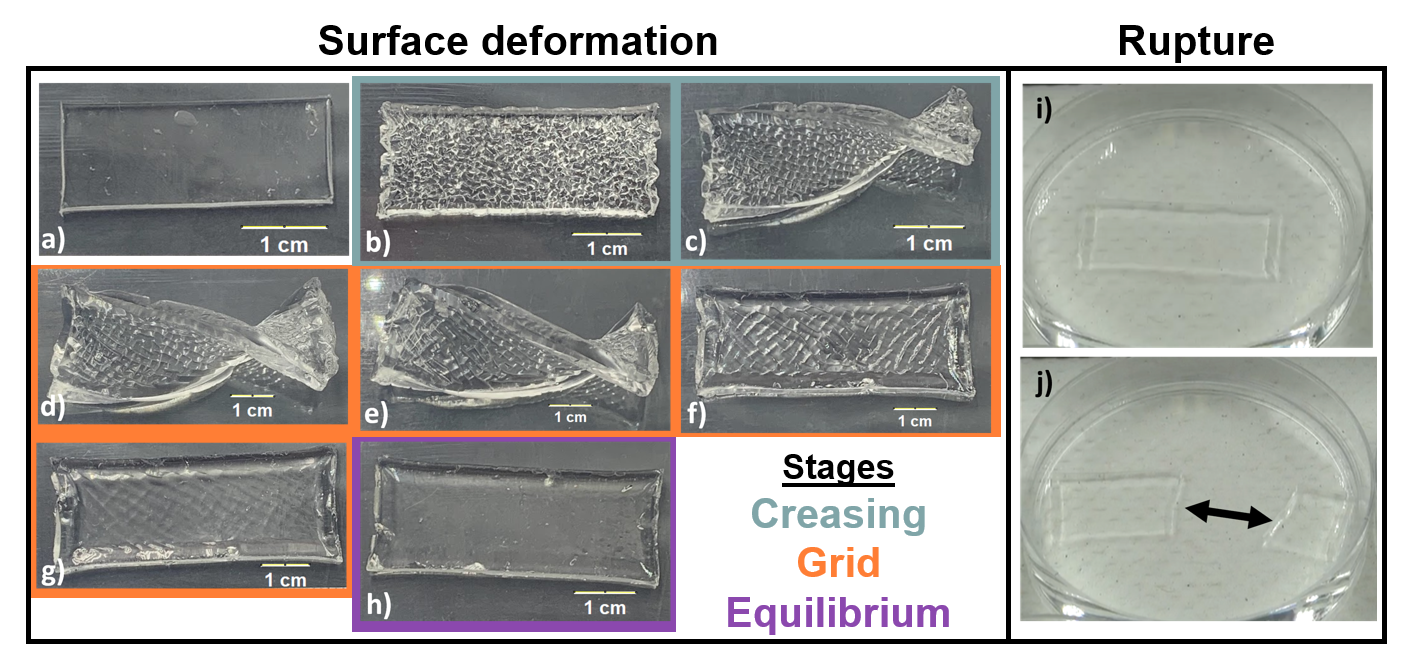}
    \caption{\textbf{Examples of instability behaviors during swelling in water.} First, the evolution of surface deformation for a 1:99, PEGDA-700 sample is shown at various swelling times, and the stages of instability are categorized as follows: a) 0 min before swelling begins, then \textcolor{PineGreen}{creasing} begins from b) 5 min, c) 10 min, then d) the \textcolor{RedOrange}{grid} pattern emerged at 20 min, e) 30 min, f) 50 min, g) 70 min, and h) the surface became smooth by 168 min, indicating \textcolor{Purple}{equilibrium}. Second, i) a 10:90, PEGDA-700 sample is shown at the start of swelling, and j) spontaneous rupture was observed after 14 minutes of swelling.
}
    \label{fig:examples}
\end{figure}

Qualitative observation of swelling behavior for each gel formulation (\textbf{Table \ref{table: formulations}}) reveals two types of instabilities during transient solvent diffusion; 
complex surface deformation (e.g., \textbf{Fig. \ref{fig:examples}}a-h) and catastrophic rupture (e.g., \textbf{Fig. \ref{fig:examples}}i,j). 
Surface deformations were observed for six formulations during swelling in water: 1:99, 5:95, and all formulations employing the 10k cross-linker (pink entries, \textbf{Table \ref{table: formulations}}). The remaining formulations (10:90, 20:80, and 40:60, gray entries, \textbf{Table \ref{table: formulations}}) did not exhibit macroscopic surface deformations during swelling in water, but often exhibited self-rupture. All samples that underwent surface deformation during transient swelling exhibited the same three-stage surface feature evolution: \textcolor{PineGreen}{creasing}, \textcolor{RedOrange}{grid}, and \textcolor{Purple}{equilibrium}. Immediately after being placed in an excess of solvent (e.g., on the order of seconds), small \textcolor{PineGreen}{creases} (detectable with the naked eye) appeared (see \textbf{Fig. \ref{fig:examples}}b). As swelling progressed (e.g., over many minutes), these surface features grew and coalesced, and additional self-contact occurred between ``cells" as seen in \textbf{Fig. \ref{fig:examples}}c,d. 
As swelling continued, a long-range \textcolor{RedOrange}{grid} pattern emerged (see \textbf{Fig. \ref{fig:examples}}e,f, and g). At later swelling stages (e.g. 70 minutes of swelling or longer), the grid was still visible, although it became less distinct. Finally, the surface of the gel became smooth again (\textbf{Fig. \ref{fig:examples}}h) once the gel reached its final, \textcolor{Purple}{equilibrium} state. Equilibrium swelling is verified by constant $Q(t)$ values (e.g., plateauing to $Q_{eq}$).

Prior experimental work describes related surface deformations, such as creasing in gels that are attached to a substrate \cite{Kim2014, Xu2013, Dupont2010, Trujillo2008} or under constraint \cite{Zalachas2013, Kashihara2022}, as well as bulk buckling in gels with an engineered gradient in cross-link density \cite{Guvendiren2009, Guvendiren2010} or layered structure\cite{Ilseng2019, Wu2013}. 
However, to the best of our knowledge, the observed ``\textcolor{RedOrange}{grid}'' surface patterning during swelling has only been seen in one previous study, which found that this instability pattern was coupled to bulk buckling of hydrogel discs\cite{Takahashi2016}. 
The similar, anisotropic grid-like crease pattern in our system 
provides perspective on how
other sources of anisotropic internal stresses can be engineered.
Self-rupture behavior is addressed in computational studies on constrained swelling leading to fracture \cite{Plummer2023}, and in experimental studies where rupture occurred during free swelling \cite{Leslie2021}, as well as pre-programmed rupture induced by non-uniform swelling \cite{DeSilva}.
Theory has been developed to understand how stress due to swelling contributes to surface deformation in free-swelling spheres \cite{Curatolo2017} and in constrained gel blankets \cite{Toh2015}, 
and how in soft materials internal stresses are often dissipated through instabilities (e.g. creases, wrinkles, or buckling).
Here, we explore how surface deformations are coupled to both material properties (network architecture) and swelling kinetics (solvent selection) in a free-swelling gel, where no mechanical constraints are applied during the swelling process.
Understanding the coupling between these swelling control parameters in an unconstrained system is necessary to design and tailor structured and dynamic material interfaces.

\subsection{\textbf{Impact of Internal Network Constraints}}
\begin{figure}[h!]
    \centering
    \includegraphics[scale=1.1]{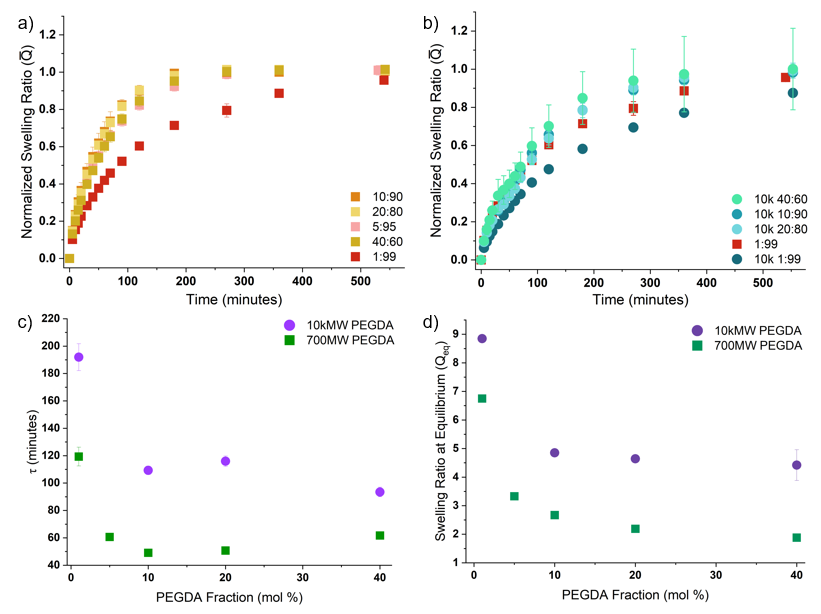}
    \caption{\textbf{Tuning internal network constraints leads to measurable differences in swelling behavior (i.e. normalized swelling ratio ($\overline{Q}$), time to equilibrium ($\tau$), and equilibrium swelling ratio ($Q_{eq}$)).}
    a) Normalized swelling ratio ($\overline{Q}$) vs.\ time, which represents the fraction of the total swelling capacity for gel formulations containing the 700MW PEGDA cross-linker. 
    b) Normalized swelling ratio ($\overline{Q}$) vs.\ time for gel formulations utilizing the 10,000MW PEGDA cross-linker. Here the 1:99 formulation with the 700MW cross-linker is also plotted for reference. 
    c) 
    $\tau$, the characteristic time parameter for $\overline{Q}$ (\textbf{Eq. \ref{tau}}), is plotted versus the fraction of cross-linker (PEGDA) in all gel formulations.
    An example fitting to obtain $\tau$ is provided in \textbf{Fig. S\ref{fig:tau fitting}}.
    Generally, these data show that a decrease in cross-linker fraction corresponds to a decrease in $\overline{Q}$ at early swelling times, as well as an increase in the total time required to reach equilibrium.
    d) The equilibrium swelling ratio ($Q_{eq}$) for all gel formulations are plotted as a function of cross-linker fraction.
    All data are for gels swelled in water.
    }
    \label{fig: NSR internal network constraints}
\end{figure}

All samples containing the high MW cross-linker (PEGDA-10,000, denoted by ``10k" in our sample notation) exhibited surface creasing and deformation during swelling in water; no rupture events were observed regardless of cross-linker fraction (mol\%). This is in contrast to formulations containing the lower molecular weight cross-linker (PEGDA-700), where  high cross-linker content (10 mol\% or greater) corresponds to a high likelihood of rupture and no observable surface creasing or deformation. 
This indicates that internal network constraints significantly impact swelling kinetics and solvent diffusion.
To quantify this behavior, we measure the swelling ratio $Q(t)$  (\textbf{Eq. \ref{eq: Q}}, 
\textbf{Fig. S\ref{tab:equilibrium sr}});
we report swelling for all samples as the normalized swelling ratio $\overline{Q}$, which is 0 in a dry gel and 1 for a fully swollen (equilibrium) gel
(\textbf{Fig. \ref{fig: NSR internal network constraints}}a,b). At each time point, 
$\overline{Q}$ indicates the fraction of the total swelling capacity the gel has obtained. Higher $\overline{Q}$ at earlier times indicate samples that quickly reach a high fraction of equilibrium swelling ratio ($Q_{eq}$), which is often a result of the total swelling capacity being relatively small (e.g., $Q_{eq}$ for the 40:60 gels is  1.88, whereas for 1:99 gels this value is  6.75 - \textbf{Fig. \ref{fig: NSR internal network constraints}}d). Thus only a small amount of water is imbibed to reach a high $\overline{Q}$. 

The influence of cross-linker MW can be understood by comparing the $\overline{Q}$ data for PEGDA-700 formulations (\textbf{Fig. \ref{fig: NSR internal network constraints}}a) with PEGDA-10,000 formulations (\textbf{Fig. \ref{fig: NSR internal network constraints}}b). Overall, we find that $\overline{Q}$ is higher at early times for the PEGDA-700 samples.
While most formulations with the PEGDA-700 cross-linker reach equilibrium relatively quickly, the 1:99 formulation takes a much longer time. 
Similar behavior is observed if the cross-linker MW is increased (10K - \textbf{Fig. \ref{fig: NSR internal network constraints}}b).  Gels with 1 mol\% of the higher MW cross-linker (10k 1:99) are slower to achieve equilibrium, and increasing cross-linker fraction generally decreases the time to equilibrium. 


To quantify the differences in time to equilibrium across various network architectures, each $\overline{Q}$ dataset was fit with the following equation:
\begin{align} \label{tau}
    \overline{Q}=1-exp(\frac{-t}{\tau})
\end{align}
where $\tau$ describes the characteristic swelling time; this exponential relationship is commonly used to describe gel swelling
\cite{Li1990, Berens1978, Zhang2020, Sievers2021, Ganji2010, Gruber2023}.
The measured $\tau$ values (\textbf{Fig. \ref{fig: NSR internal network constraints}}c) decrease with increasing cross-linker fraction, regardless of the cross-linker employed. When PEGDA-700 is employed, the 1:99 formulation experiences a much larger time to reach equilibrium ($\tau\simeq$120 minutes) than the formulations with higher PEGDA-700 content (all $\tau\leq$61 minutes).
Similarly, when PEGDA-10,000 is used as the cross-linker, the 10k 1:99 formulation exhibits the highest $\tau$ value ($\tau\simeq$190 minutes), and increasing cross-linker fraction lowers $\tau$.
Overall, networks with lower cross-linker content imbibe a higher fraction of water (see \textbf{Fig. \ref{fig: NSR internal network constraints}}d), and also take more time to reach equilibrium.
In other words, when the constraints in the network are increased (via increased cross-linking), the total capacity to swell decreases and therefore less solvent is required to reach equilibrium.

At all cross-linker loadings, the formulations using the PEGDA-10,000 cross-linker have a larger $\tau$,
which demonstrates that increasing the length of the cross-linker (PEGDA) corresponds to an increase in the time to reach equilibrium. 
The increased  flexibility afforded by a higher MW cross-linker (PEGDA-10,000)
means that the network has more freedom to 
accommodate solvent during swelling. 

\subsection{\textbf{Effect of External Diffusive Pressure on Swelling}}
The $\overline{Q}$ and $\tau$ behavior for swelling PEGDA-700 and PEGDA-10,000 gels in water highlight how the rate of solvent imbibement during swelling is influenced by the internal constraints of the polymer network.
This is consistent with the Flory-Rehner (FR) polymer solution thermodynamics model, which postulates that gel swelling is described by an overall osmotic pressure ($\Pi$), that is composed of two main terms, the pressure of mixing ($\Pi_{mix}$) and the pressure of elasticity ($\Pi_{el}$) (see \textbf{Eq. \ref{osmotic pressure}}) \cite{Louf2021, Quesada-Perez2011, Flory, Sakai, Hirotsu1994, Rubinstein}.
The sum of the pressure of mixing ($\Pi_{mix}$) and the pressure of elasticity ($\Pi_{el}$) equals the overall osmotic pressure ($\Pi$) that the material experiences during swelling.
Total osmotic pressure ($\Pi$) and mixing pressure ($\Pi_{mix}$) are positive values that represent the forces that \textit{drive} swelling, while ($\Pi_{el}$) is always negative, 
representing the \textit{resistance} to swelling that the elasticity of the polymer network asserts.
\begin{align} \label{osmotic pressure}
    \Pi_{mix} = -\frac{k_B T}{\alpha^3}\left[\phi + \ln{\left(1-\phi\right)} + \chi\phi^2\right] \\
    \Pi_{el} = \frac{k_B T N_c}{V_0}\left[\frac{\phi}{2\phi_0}-\left(\frac{\phi}{\phi_0}\right)^{\frac{1}{3}}\right] \nonumber \\ 
    \Pi = \Pi_{mix} + \Pi_{el} \nonumber
\end{align}
$\Pi_{el}$ accounts for network constraints through the term $N_c$, which is a count of all chains within the network\cite{Flory1943swelling}, and is typically on the order of $10^{18} - 10^{20}$ and increases with cross-link density. 
When cross-linking increases (e.g., increase in $N_c$), $\Pi_{el}$ will become more negative, and there is more resistance to swelling. In other words, increasing cross-linking decreases the freedom chains have to accommodate imbibed solvent. 

\begin{table}[ht]
    \centering
    \caption{\textbf{Definition of terms} in Flory-Rehner's theory of osmotic pressure}
    \begin{tabular}{|c|c|}
    \hline
        Term & Definition \\
        \hline
        $k_B$ & Boltzmann's constant \\
        \hline
        $\alpha$ & Effective solvent diameter \\
        \hline
        $\phi$ & Volume fraction of polymer \\
        \hline
        $\phi_0$ & Initial volume fraction (equals 1 for neat polymer) \\
        \hline
        $\chi$ & Flory-Huggins interaction parameter \\
        \hline
        $N_c$ & Number of chains in the polymer network \\
        \hline
        $V_0$ & Initial volume \\
        \hline
    \end{tabular}
    \label{tab:definitions}
\end{table}

$\Pi_{mix}$ represents the driving force for swelling that originates from the attraction between the polymer network and the solvent, which is often characterized by the interaction parameter $\chi$. $\Pi_{mix}$ also accounts for how the size of a solvent molecule impacts physical mixing.
Inspired by the equilibrium FR model, we believe that both the external diffusive pressure during swelling and the elastic network both can be leveraged, even in the transient state, to manipulate the rate of solvent imbibement.

The experiments presented up to this point explicitly manipulate the elastic contribution to swelling (i.e., network constraints). 
To systematically vary the external diffusive pressure and establish how that impacts instability behaviors, PEGDA-700 samples were swelled in varying (co)solvents comprised of water and ethanol. 
Initial swelling rates, which were estimated by performing a linear fit to the first 30 minutes of the swelling ratio $Q(t)$, were calculated for all gel formulations with PEGDA-700 as the cross-linker, and are plotted in \textbf{Fig. S\ref{fig: rate in water and ethanol}}.
Overall, swelling in ethanol resulted in a significant decrease in initial swelling rate at all PEGDA-700 fractions.
Additionally, for swelling in both water and ethanol, the initial rate of swelling decreases as cross-linker fraction increases. This data demonstrates how the external diffusive pressure, which can be manipulated via solvent selection, is a means to further manipulate the timescale of swelling.


\begin{figure}[h!]
    \centering
    \includegraphics[scale=0.9]{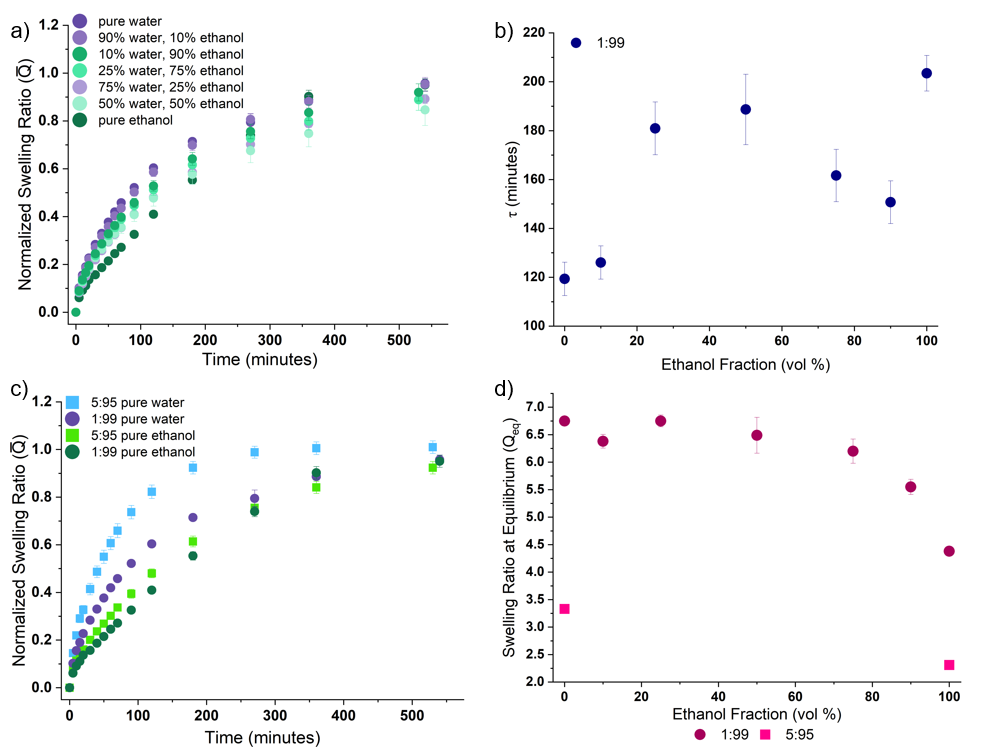}
    \caption{\textbf{Swelling kinetics for ethanol, water, and ethanol/water (co)solvents.}
    a) Evolution of the normalized swelling ratio ($\overline{Q}$) for the 1:99, PEGDA-700 formulation in a series of water:ethanol co-solvents. 
    b) The characteristic swelling time, $\tau$, for the 1:99, 700MW formulation for all (co)solvents. 
    c) Comparison of the evolution of the normalized swelling ratio ($\overline{Q}$)  for the 1:99 and 5:95 PEGDA-700 formulations in pure water and pure ethanol. 
    d) Equilibrium swelling ratio ($Q_{eq}$) for the 1:99, PEGDA-700 formulation for all (co)solvents shows that increasing the amount of ethanol in the (co)solvent generally decreases the $Q_{eq}$.
    The $Q_{eq}$ for 5:95, PEGDA-700 formulation in both pure water and pure ethanol are included for reference, and highlight the impact of cross-linker fraction on $Q_{eq}$.
    }
    \label{fig: NSR external diffusive pressure}
\end{figure}

To better understand the impact of external diffusive pressure, particularly on transient surface instabilities, the PEGDA-700 1:99 formulation was chosen for further analysis.
Specifically, $\overline{Q}$ was characterized for the 1:99 formulation when swelling was conducted in pure water, pure ethanol, and 5 co-solvent mixtures (\textbf{Fig. \ref{fig: NSR external diffusive pressure}}a). 
The solvent-driven changes in $\overline{Q}$ are consistent with the initial swelling rate and are further reflected in the $\tau$ parameters (\textbf{Fig. \ref{fig: NSR external diffusive pressure}}b).
To explore the effect of changing network architecture, we measured swelling for two different cross-link densities; 1:99 and 5:95.
As shown in \textbf{Fig. \ref{fig: NSR internal network constraints}}b, increasing cross-link density leads to faster equilibration times. This may be attributed to the fact that the network imbibes less overall solvent ($Q_{eq}$, \textbf{Fig. \ref{fig: NSR external diffusive pressure}}d). 

Water is known to be a better solvent for these PEG gels, and thus we expect that as the fraction of ethanol in the (co)solvent mixture increases, $\overline{Q}$ will decrease and $\tau$ will increase.
Surprisingly, we find instead that 
$\tau$ changes non-monotonically with ethanol vol\% (\textbf{Fig. \ref{fig: NSR external diffusive pressure}}b).
This indicates that there are competing interactions between water and ethanol that interfere with diffusion into the polymer network during swelling in a mixed solvent.

To understand this non-monotonic relationship, we must quantify the attraction between the (co)solvent molecules and the polymer network.
The standard method of quantifying interactions is the Flory-Huggins (FH) interaction parameter ($\chi$). 
$\chi$ can be estimated between two species ($\alpha$ and $\beta$) using \textbf{Eq. \ref{FH interaction parameter}} \cite{Dudowicz2015}:
\begin{align}\label{FH interaction parameter}
    \chi_{\alpha\beta}=(\frac{z}{2})\frac{[\epsilon_{\alpha\alpha}+\epsilon_{\beta\beta}-2\epsilon_{\alpha\beta}]}{k_BT}\\
    \alpha \equiv A,B, \beta \equiv B, C \neq \alpha \nonumber
\end{align}
Here, $z$ is the coordination number (i.e., the total number of molecules bonded to a central atom), $\epsilon_{\alpha\alpha}$ is the self-attractive energy for species $\alpha$, $\epsilon_{\beta\beta}$ is the self-attractive energy for species $\beta$, and $\epsilon_{\alpha\beta}$ is the interaction energy between species $\alpha$ and $\beta$. 
Positive (+) values of $\chi$ indicate that mixing is not energetically favorable, whereas negative (-) values of $\chi$ indicate that mixing is favorable. 
Experimentally determined $\chi$ for a ternary mixture of water, ethanol, and linear PEG chains with molecular weights 8650 $\frac{g}{mol}$ and 12600 $\frac{g}{mol}$ are reported in \textbf{Table \ref{tab:water-PEG-ethanol interactions}} \cite{Luh1988}. 
While these measurements are for linear PEG systems,, and may not directly correspond to $\chi$ measured for a cross-linked PEG system, they can help us understand how our ternary mixture interacts.

\begin{table}[h!]
\centering
\caption{\textbf{Experimentally measured Flory-Huggins interaction parameters ($\chi$).}
    $\chi$ for the ternary water-PEG-ethanol mixture were determined by Song-Ping Luh for two molecular weights of PEG \cite{Luh1988}.}
\begin{tabular}{|c|c|c|}
   \hline
    Flory-Huggins & & \\
    interaction parameter & PEG 8650 & PEG 12600 \\
    \hline
    $\chi_{water-PEG}$ & 1.0001 & 0.6318 \\
    \hline
    $\chi_{water-ethanol}$ & -4.4412 & -4.2339 \\
    \hline
    $\chi_{PEG-ethanol}$ & -0.6862 & -0.7451 \\
    \hline
\end{tabular}
    \label{tab:water-PEG-ethanol interactions}
\end{table}

\textbf{Table \ref{tab:water-PEG-ethanol interactions} }shows that mixing of water and PEG is slightly unfavorable, mixing PEG and ethanol is favorable, and mixing water and ethanol is significantly favorable (due to the large, negative $\chi_{water-ethanol}$). 
This can explain the non-monotonic swelling behavior we observe in (co)solvent mixtures: 
water and ethanol are significantly attracted to one another, meaning the interaction between these solvent species likely competes with the driving force of swelling \cite{Dudowicz2015}.
The normalized swelling ratio ($\overline{Q}$) data for swelling in (co)solvent mixtures supports this finding (\textbf{Fig. \ref{fig: NSR external diffusive pressure}}a). 
The $\overline{Q}$ behavior for 10/90 and 90/10 (vol/vol, water/ethanol) (co)solvents are very similar to pure water behavior (\textbf{Fig. \ref{fig: NSR external diffusive pressure}}a),
which is similar to findings from ternary mixtures with linear PEG polymers\cite{Dudowicz2015}. More specifically, when (co)solvent species experience significant interaction with each other, the miscibility between the (co)solvent mixture and the polymer network is decreased, and immiscibility reaches its maximum when the volume ratio between the (co)solvent species is equal to one.
Indeed, \textbf{Fig.\ref{fig: NSR external diffusive pressure}}b shows that the 50/50 (vol/vol) (co)solvent mixture exhibits the highest $\tau$ value, indicating that the ternary mixture reaches its peak immiscibility when water and ethanol are mixed in equal parts.
This is particularly interesting, as this behavior does not follow established co-non-solvency and co-solvency behaviors in gel swelling\cite{mukherji_collapse_2018,higaki_cononsolvency-induced_2023,steinbeck_porous_2024}, as our two (co)solvents have differing interactions with the gel (e.g. positive vs. negative $\chi$).

While the $\chi$ values for water-PEG and PEG-ethanol indicate better mixing when gels are swollen in ethanol, our data shows reduced $Q_{eq}$ and initial swelling rate in ethanol (\textbf{Fig. \ref{fig: NSR external diffusive pressure}}d and \textbf{Fig. S\ref{fig: rate in water and ethanol}}, respectively). This can be explained by considering the solvent molecular size. 
Flory-Rehner theory (see \textbf{Eq. \ref{osmotic pressure}}) states that the pressure of mixing ($\Pi_{mix}$) is inversely proportional to the solvent diameter cubed, meaning that increasing the size of the solvent molecule would significantly decrease the pressure of mixing (i.e., the driving force for swelling). The effective diameter of a water molecule is 2.75 $\mathring{A}$ \cite{Schatzberg1967}, and the diameter of an ethanol molecule is 4.5 $\mathring{A}$ \cite{Song2015}. This is consistent with our experimental observation that the network swells considerably more in water. Although ethanol is a better solvent for \textit{linear} PEG chains, its larger size hinders its ability to mix with the three-dimensional, cross-linked PEG network.

\subsection{\textbf{Examining Surface Creasing during Swelling}}
\begin{figure}[h!]
    \centering
    \includegraphics[width=0.65\textwidth]{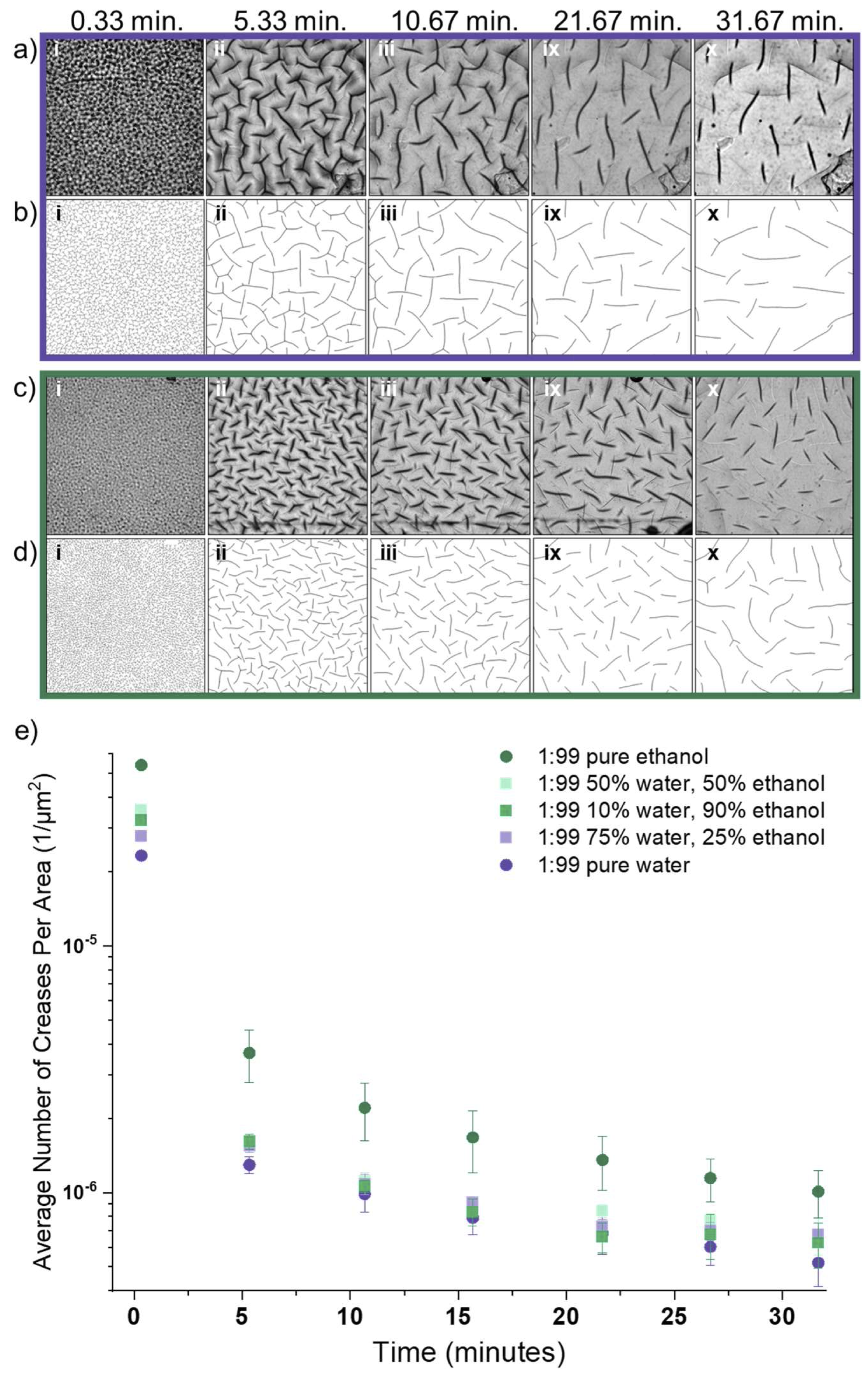}
    \caption{\textbf{Surface creasing during swelling in water, ethanol, and water-ethanol (co)solvents.}
    \textbf{a)} Microscope images taken of a 1:99, PEGDA-700 sample swelling in \textit{water} for i) 0.33, ii) 5.33, iii) 10.67, ix) 21.67, and x) 31.67 minutes.
    \textbf{b)} The creases in each microscope image were manually traced in order to accurately identify and binarize the patterns, 
    \textbf{c)} Microscope images of a 1:99, PEGDA-700 sample swelled in \textit{ethanol} for i) 0.33, ii) 5.33, iii) 10.67, ix) 21.67, and x) 31.67 minutes, as well as the 
    \textbf{d)} corresponding traced crease patterns.
    \textbf{e)} The average number of creases was measured and divided by the image area for 1:99, PEGDA-700 formulation samples swelled in pure water, pure ethanol, and three (co)solvent mixtures. This analysis reveals there are a large number of creases at the onset of swelling, and a significant decrease in number of creases after five minutes of swelling. The number of creases continue to gradually decrease over 30 minutes of observation.
    }
    \label{fig:avg numb of creases}
\end{figure}

Changing the solvent mixture changes the rate of solvent imbibement, and thus impacts transient material stress.  To quantify these changes, we studied the evolution of surface instabilities during swelling in water, ethanol, and (co)solvent mixtures.  We used light microscopy to image surface instabilities \textit{in situ} and characterize pattern evolution.
The surface deformation of the 1:99, PEGDA-700 formulation was 
characterized during swelling in ethanol, water, and (co)solvent mixtures (\textbf{Fig. \ref{fig:avg numb of creases}}). 

The observed crease evolution in water (\textbf{Fig. \ref{fig:avg numb of creases}}a, corresponding binary tracings \textbf{Fig. \ref{fig:avg numb of creases}}b) and in ethanol (\textbf{Fig. \ref{fig:avg numb of creases}}c, corresponding binary tracings \textbf{Fig. \ref{fig:avg numb of creases}}d),
show that surface deformation began rapidly once the gel was placed in solvent.
We note that the images show creasing on both the bottom (in-focus sharp lines) and top of the sample (larger darker out-of-focus lines); Analysis was done on the creases on the bottom of the sample as highlighted in \textbf{Fig. S\ref{fig:crease highlights water}} and \textbf{Fig. S\ref{fig:crease highlights ethanol}}.
\textbf{Fig. \ref{fig:avg numb of creases}} shows that swelling in ethanol leads to a higher number of creases when compared to swelling in water.
This is quantified in \textbf{Fig. \ref{fig:avg numb of creases}}e, which shows the mean density of creases as a function of swelling time  in the various (co)solvent mixtures. 
The density of creases rapidly decreases in time.
This large, relatively fast decrease is due to relaxation of the outermost layer of the gel as diffusion of the solvent continues deeper into the network.
At all time points, 
swelling in pure ethanol results in significantly more creases than any other (co)solvent, and swelling in pure water shows the lowest number of creases. The three co-solvent mixtures explored fell between these extremes and are statistically similar.

As shown in \textbf{Fig. S\ref{fig: rate in water and ethanol}}, swelling the 1:99 formulation in water significantly increases the initial rate of swelling compared to ethanol. The slower imbibement of ethanol corresponds to a lower driving force for swelling and less mixing pressure ($\Pi_{mix}$) forcing the network to relax and dissipate the compressive stress at the surface.
Despite the fact that the network imbibes significantly less fluid during swelling in ethanol as compared to water, this results in twice the number of surface creases (\textbf{Fig. \ref{fig:avg numb of creases}}e). 
This suggests that (co)solvent selection and the associated diffusivity within a gel network is a significant parameter when tailoring the scale of instabilities that form.

Wrinkling of soft materials under compression is well-characterized, and theories have been developed to connect wrinkling wavelength directly with material properties\cite{Genzer2006}.  The features we observe in this gel system are creases (as opposed to wrinkles), as evidenced by their random orientation. This indicates the material has made self-contact, which presents a challenge for connecting these patterns to network material properties.
However, the assumption from classical wrinkling of a skin layer atop a foundation is similar to the initial states of swelling when just a thin, outermost layer of the gel has imbibed solvent.
Wrinkling theory predicts that the compressive force at the surface is related to the skin thickness, the elastic modulus and Poisson's ratio of both the skin and the bulk \cite{Genzer2006}. As the compressive force increases, the wavelength of wrinkles decreases. 
Thus, the lower number of features (\textbf{Fig.\ref{fig:avg numb of creases}}e), and 
the higher spacing (i.e., ``wavelength") we observe when swelling in water likely represents lower compressive forces at the surface, particularly at early time points.

\begin{figure}[h!]
    \centering
    \includegraphics[width = 12 cm]{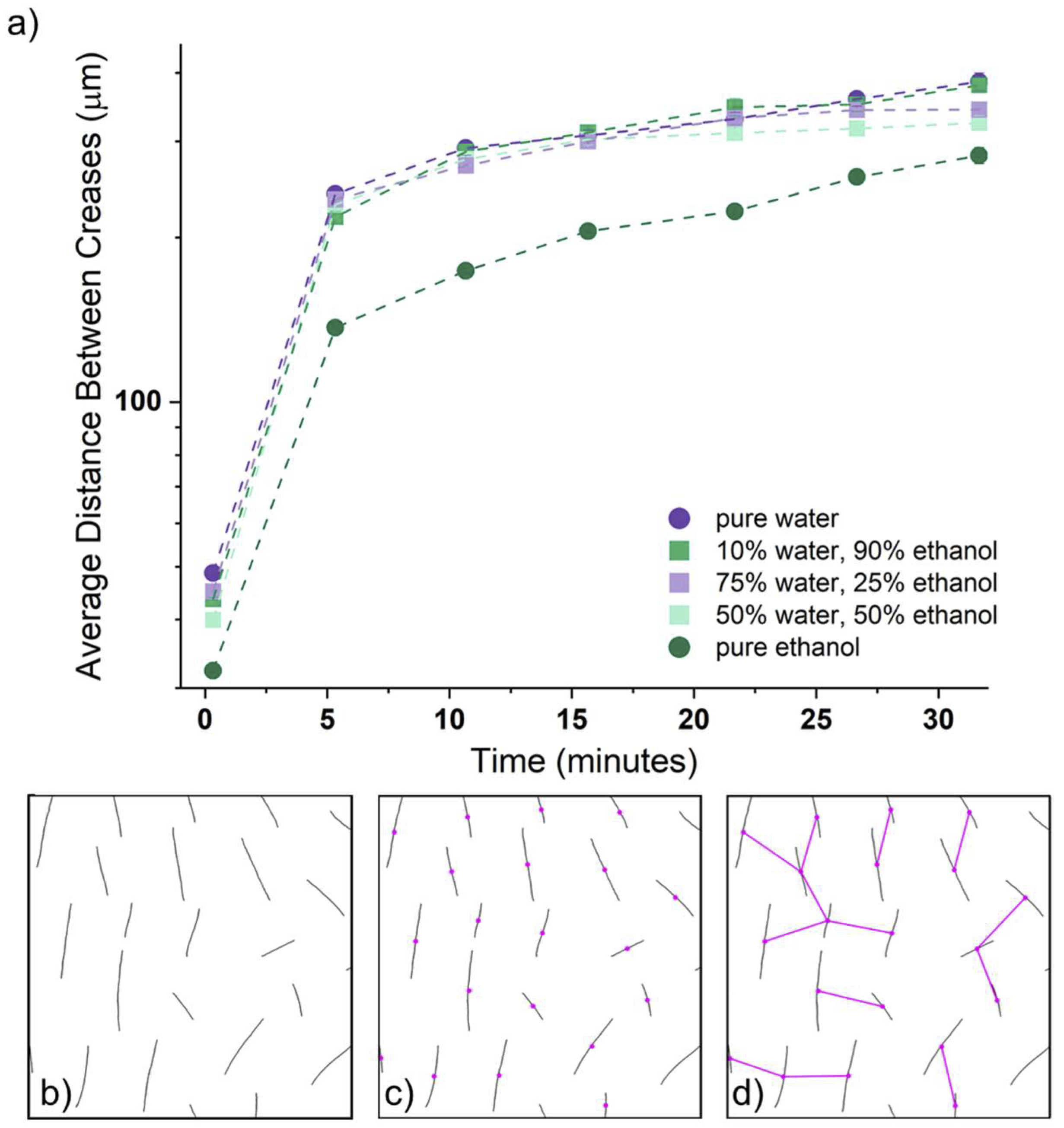}
    \caption{\textbf{Characteristic distance between creases.}
    a) Analysis of the traced crease patterns was performed to determine the average distance between creases for the 1:99, PEGDA-700 formulation swelled in pure water, pure ethanol, and three (co)solvent mixtures. At all times, the distance between creases in pure ethanol is smaller than the other conditions.
    b) Example tracing of swelling in water to demonstrate how the average distance was determined. First, each individual crease was identified,
    c) followed by calculation of the geometric mean, and a location was assigned for each crease (indicated by purple dots).
    d) For each crease, the closest neighbor was identified, and the distance between them was measured. This process was repeated for each crease, and any repeated distances (i.e., a pair of creases that reported one another as their nearest neighbor with the same distance value) were eliminated. 
    (a) reports the mean distance across the triplicate samples for each solvent condition.
    }
    \label{fig: avg dist btwn creases}
\end{figure}

To quantify crease spacing, we measure the mean inter-crease distance for each solvent condition (\textbf{Fig. \ref{fig: avg dist btwn creases}}a). 
This analysis shows that the average crease spacing at all times is lowest for ethanol, indicating higher compressive stress. As time evolves, the crease distance rapidly increases, which correlates to a relaxation of the polymer network with time.
Similar to the average number of creases, all solvent conditions show a dramatic increase in the distance between creases between 0.33 min. and 5.33 min. (i.e., between a 300\% and a 400\% increase). 
This increase in distance between creases (our analogue to wavelength) indicates a decrease in compressive stress at the surface due to the immediate relaxation that occurs at the surface of the gel.

Interestingly, at late times, the 50/50 
water/ethanol (co)solvent plateaued and diverged slightly from the grouping of the pure water and other (co)solvent mixtures.
A similar trend appeared in the number of creases 
(\textbf{Fig. \ref{fig:avg numb of creases}}a).
This result can be understood in the context of the (co)solvent theory that immiscibility is greatest when the ratio $\frac{\phi_{ethanol}}{\phi_{water}}$ is equal to 1. 
The interaction between water and ethanol that competes with swelling is greatest when the (co)solvent contains equal parts of water and ethanol. This significant interference leads to the (co)solvent mixture behaving as a worse solvent.

\newpage
\section{\textbf{Conclusions
}}
We explore the impact of internal network constraints (cross-linking) and external diffusive pressure (solvent quality) on swelling kinetics and surface deformation dynamics in PEG-based gels. 
We find that networks with fewer internal constraints (which have a higher overall capacity for swelling, $Q_{eq}$), experience much longer transient swelling.
Increasing the molecular weight between cross-links decreases the elastic pressure, and thus increases the time it takes to reach equilibrium. 
We find that ethanol, despite having a large negative $\chi$, is a worse solvent for the network than water. This demonstrates that while imbibement is a function of chemical affinity ($\chi$), this competes with the effect of molecular size.  Swelling experiments in water-ethanol (co)solvent mixtures reveal that the strong affinity between water and ethanol competes with the driving force of swelling.

Measurements of surface instabilities show that immediately after being placed in a solvent, densely packed creases rapidly appear on the surface of the gel.  The inter-crease spacing increases in time as diffusion progresses and the material begins to relax. Furthermore, we find that swelling in ethanol results in a higher crease density, which likely indicates that larger compressive stresses are present.
Thus, swelling in ethanol not only slows down swelling kinetics, but it also alters the dynamics of surface deformations.
This reveals that the driving force for mixing is lower for swelling in ethanol as compared to water, and therefore the hydrogel network experiences less pressure to relax and dissipate the stress that builds due to swelling and solvent diffusion. 
The physics involved in generating macroscopic surface deformations is complex, and thus it is challenging to use the geometry of these features to directly quantify material stresses.  However, this work establishes a path towards understanding the complex dynamics at play during transient swelling and offers a new way to quantify material response.


\newpage
\section*{Acknowledgements}
During this study, Alyssa VanZanten was fully supported and Shih-Yuan Chen was partially supported by NSF Grant DMR 2311697.
The authors would like to thank our labmates Sabrina Curley, Denghao Fu, and Samira Kahn, for their support and rigorous discussion.
\newpage
\singlespacing
\bibliographystyle{unsrt}
\bibliography{bibliography}
\newpage
\setcounter{figure}{0}
\setcounter{table}{0}
\section*{Supporting Information}

\begin{table}[h!]
    \centering
        \caption{\textbf{Swelling ratio at equilibrium.} The average swelling ratio at equilibrium for all hydrogel formulations and solvent conditions.}
    \begin{tabular}{|c|c|c|}
        \hline
        Formulation & Solvent & Swelling Ratio\\
         & & at Equilibrium \\
         \hline
         1:99 & 	Water & 6.75\\
         \hline
         5:95 & 	Water & 3.33\\
         \hline
         10:90 & Water & 2.67\\
         \hline
         20:80 & Water & 2.19\\
         \hline
         40:60 & Water & 1.88\\
         \hline
         10k 1:99 &  Water & 8.85\\
         \hline
         10k 10:90 &  Water & 4.85\\
         \hline
         10k 20:80 &  Water & 4.64\\
         \hline
         10k 40:60  &  Water & 4.42\\
         \hline
         \hline
         1:99 & Ethanol & 4.38\\
         \hline
         5:95 & Ethanol & 2.31\\
         \hline
         10:90 & Ethanol & 2.13\\
         \hline
         20:80 & Ethanol & 1.77\\
         \hline
         40:60 & Ethanol & 1.79\\
         \hline
         \hline
         1:99 & 90\% water, 10\% ethanol & 6.38\\
         \hline
         1:99 & 75\% water, 25\% ethanol & 6.75\\
         \hline
         1:99 & 50\% water, 50\% ethanol & 6.49\\
         \hline
         1:99 & 25\% water, 75\% ethanol & 6.20\\
         \hline
         1:99 & 10\% water, 90\% ethanol & 5.55\\
         \hline
    \end{tabular}
    \label{tab:equilibrium sr}
\end{table}


\begin{figure}[ht]
    \centering
    \includegraphics[scale=0.55]{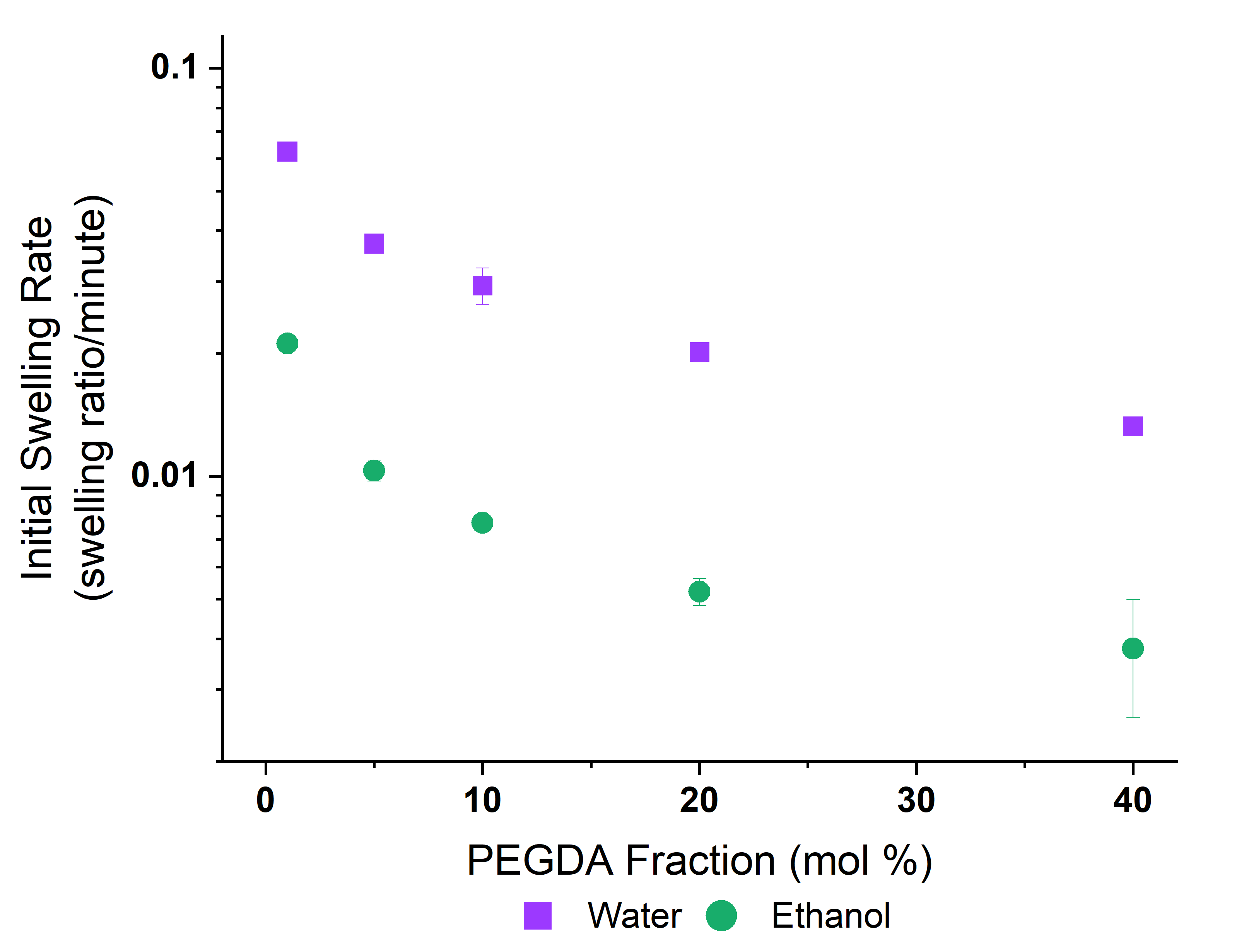}
    \caption{
    T\textbf{he initial rate of swelling in water (squares) and ethanol (circles) for gel formulations with the 700MW PEGDA cross-linker are compared.} The initial rate was measured over the first 30 minutes of swelling. This time period captures the complete evolution of observed creasing instabilities.}
    \label{fig: rate in water and ethanol}
\end{figure}

\begin{figure}
    \centering
    \includegraphics[width=0.9\textwidth]{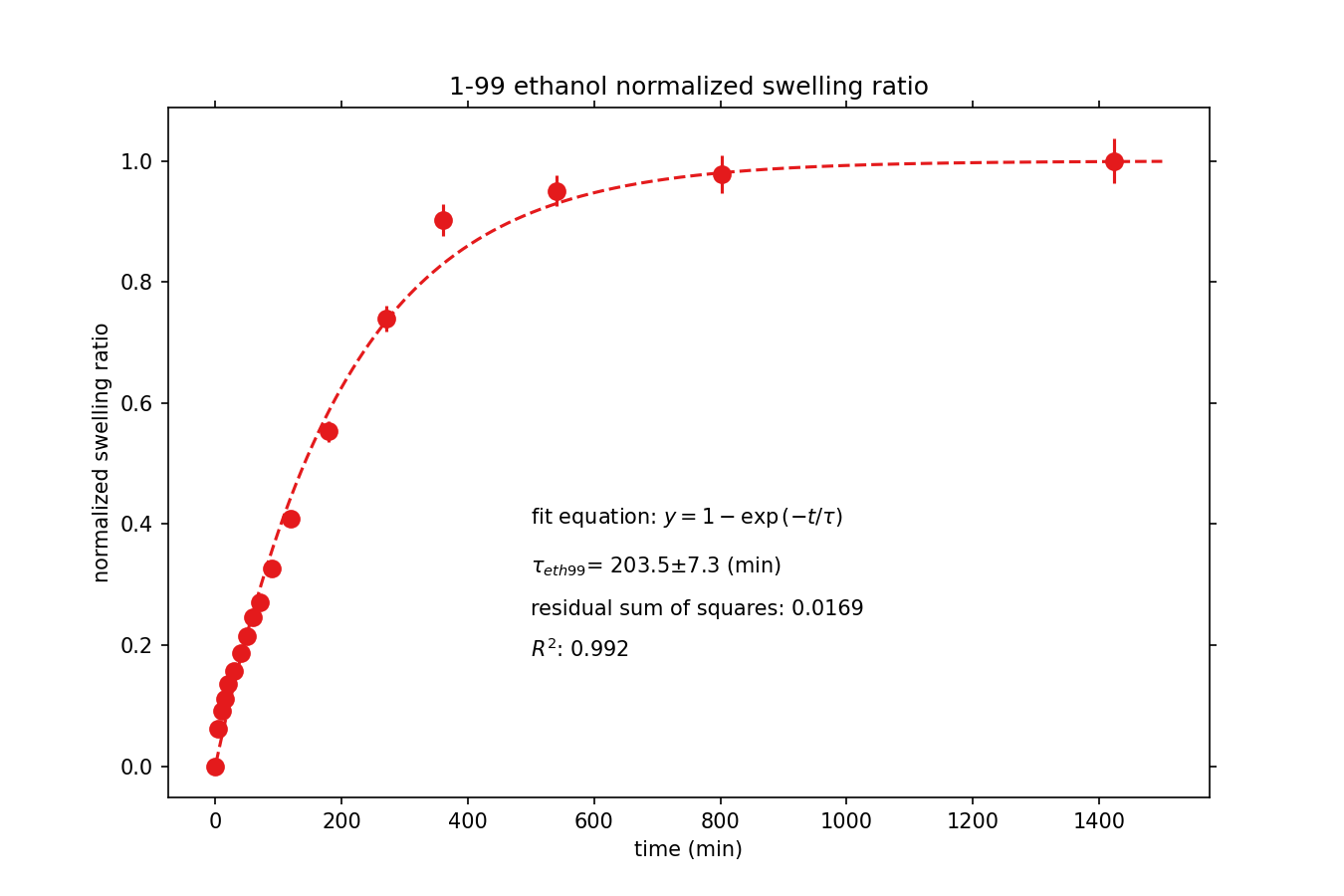}
    \caption{\textbf{Example of fitting performed to extract $\tau$, the characteristic time to equilibrium.} The normalized swelling ratio, $\overline{Q}$, is plotted and fit with an exponential equation, shown inset in this figure. This fitting output the denominator of the exponential as $\tau$, which describes the time the system, in this case 1:99, PEGDA-700 swelling in ethanol, takes to reach 64\% of the total swelling capacity.}
    \label{fig:tau fitting}
\end{figure}

\begin{figure}
    \centering
    \includegraphics[width=0.95\textwidth]{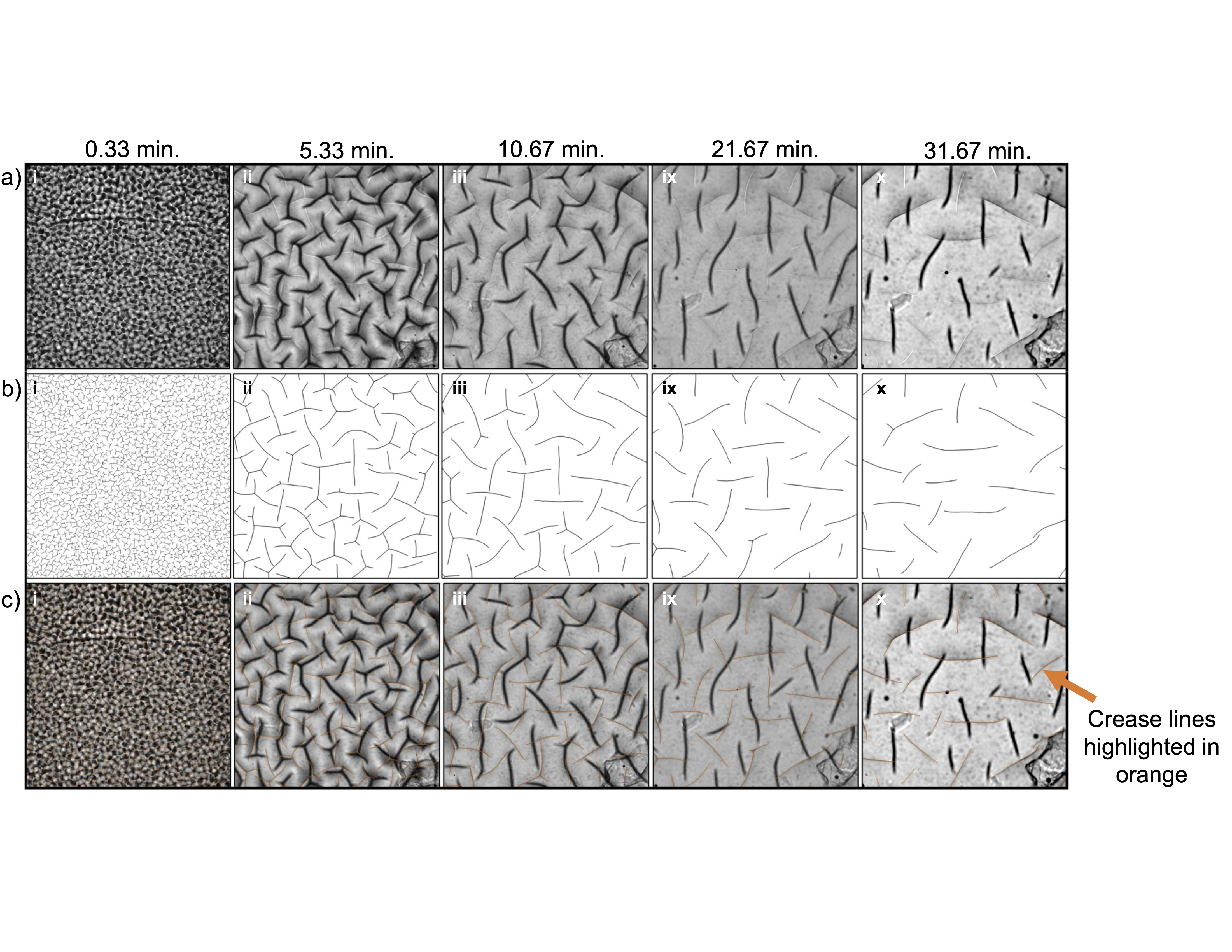}
    \caption{\textbf{Illustrating crease lines on in-focus surface during swelling in water}. a) The microscope images and b) crease tracings shown in \textbf{Fig.\ref{fig:avg numb of creases}} are repeated here. c) The traced crease lines in orange are then overlaid on top of the corresponding microscope images to highlight the location of the deformation features of interest.}
    \label{fig:crease highlights water}
\end{figure}

\begin{figure}
    \centering
    \includegraphics[width=0.95\textwidth]{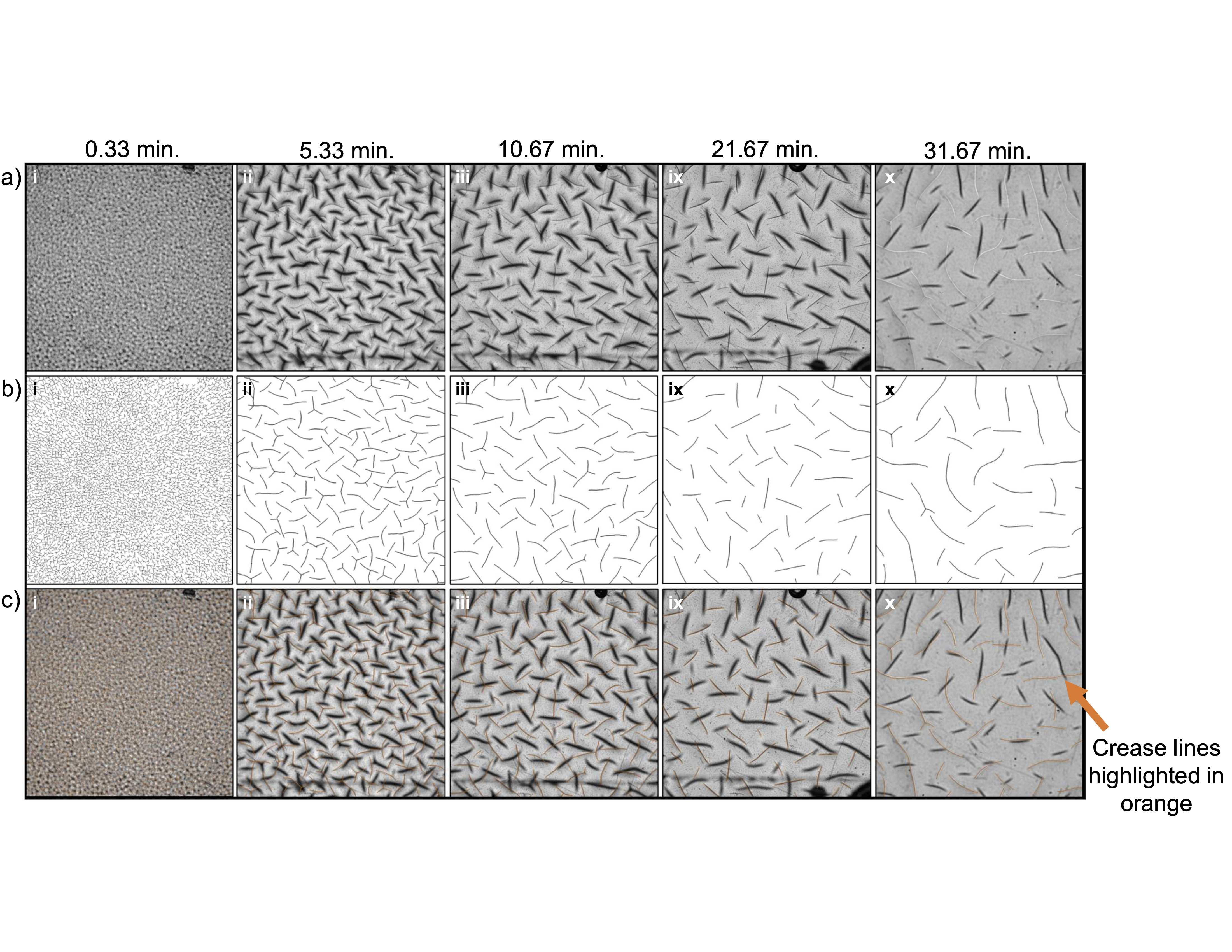}
    \caption{\textbf{Illustrating crease lines on in-focus surface during swelling in ethanol}.a) The microscope images and b) crease tracings shown in \textbf{Fig.\ref{fig:avg numb of creases}} are repeated here. c) The traced crease lines in orange are then overlaid on top of the corresponding microscope images to highlight the location of the deformation features of interest.}
    \label{fig:crease highlights ethanol}
\end{figure}
\end{document}